\documentclass[useAMS,usenatbib,usegraphicx]{mn2e}
\usepackage{url}
\usepackage{hyperref}
\usepackage{stmaryrd}

\def\aj{AJ}%
\def\apj{ApJ}%
\def\apjl{ApJ}%
\def\apjs{ApJS}%
\def\aap{A\&A}%
\def\aaps{A\&AS}%
\def\mnras{MNRAS}%
\def\hi{H\,{\sc i}}
\def\Nwhisp{339}
\def\Nugc{8147}

\begin{document}

\title[WHISP Merger Fraction \& Rate]{Quantified \hi \ Morphology IV:\\ The Merger Fraction and Rate in WHISP}

\author[B.W. Holwerda]{B. W. Holwerda$^{1,2}$\thanks{E-mail:
benne.holwerda@esa.int}, N. Pirzkal,$^{3}$ W.J.G. de Blok,$^{2}$ A. Bouchard,$^{4}$ S-L. Blyth,$^{3}$ 
\newauthor
and K. S. van der Heyden$^{2}$\\
$^{1}$ European Space Agency, ESTEC, Keplerlaan 1, 2200 AG, Noordwijk, the Netherlands\\
$^{2}$ Astrophysics, Cosmology and Gravity Centre ($ACGC$), \\  
Astronomy Department, University of Cape Town, Private Bag X3, 7700 Rondebosch, Republic of South Africa\\
$^{3}$ Space Telescope Science Institute, Baltimore, MD 21218, USA\\
$^{4}$ Department of Physics, Rutherford Physics Building, McGill University, 3600 University Street, Montreal, Quebec, H3A 2T8, Canada}

\date{Accepted 1988 December 15. Received 1988 December 14; in original form 1988 October 11}

\pagerange{\pageref{firstpage}--\pageref{lastpage}} \pubyear{2002}

\maketitle

\label{firstpage}

\begin{abstract}
The morphology of the atomic hydrogen (\hi) disk of a spiral galaxy is the first component to be disturbed by a gravitational interaction such as a merger between two galaxies. We use a simple parametrisation of the morphology of \hi \ column density maps of Westerbork \hi \ Spiral Project (WHISP) to select those galaxies that are likely undergoing a significant interaction. Merging galaxies occupy a particular part of parameter space defined by Asymmetry (A), the relative contribution of the 20\% brightest pixels to  the second order moment of the column density map ($M_{20}$) and the distribution of the second order moment over all the pixels ($G_M$). 

Based on their \hi \ morphology, we find that 13 \% of the WHISP galaxies are in an interaction (Concentration-$M_{20}$) and only 7\% based on close companions in the data-cube. This apparent discrepancy can be attributed to the difference in visibility time scales: mergers are identifiable as close pairs for 0.5 Gyr but $\sim$ 1 Gyr by their disturbed \hi \ morphology. Expressed as volume merger rates, the two estimates agree very well: 7 and $6.8 \times 10^{-3}$ mergers Gyr$^{-1}$ Mpc$^{-3}$ for paired and morphologically disturbed \hi \ disks respectively.

The consistency of our merger fractions to those published for bigger surveys such as the Sloan Digital Sky Survey, shows that \hi \ morphology can be a very viable way to identify mergers in large \hi \ survey. The relatively high value for the volume merger rate may be a bias in the selection or WHISP volume. The expected boon in high-resolution \hi \ data by the planned MeerKAT, ASKAP and WSRT/APERTIF radio observatories will reveal the importance of mergers in the local Universe and, with the advent of SKA, over cosmic times. 

\end{abstract}


\section{\label{s:intro}Introduction}

Mergers of galaxies is a driving factor in their evolution over cosmic times. Several schemes to identify merging galaxy pairs have been developed in the past decade, many based on the number of physically close pairs (in both sky coordinates as well as redshift) or on the characterization of the disturbed appearance of galaxies due to gravitational interaction \citep[often through visual inspection, e.g.,][]{Arp73b,Vorontsov-Velyaminov01,Darg09}. Both these techniques have been used to determine the interaction fraction in the local universe as well as out to high redshift in Hubble images. Using N-body simulations, one can determine how long a merger will be identified as such by both techniques; the galaxy pair is close enough, the galaxies look sufficiently disturbed. 

\begin{figure}
\centering
\includegraphics[width=0.5\textwidth]{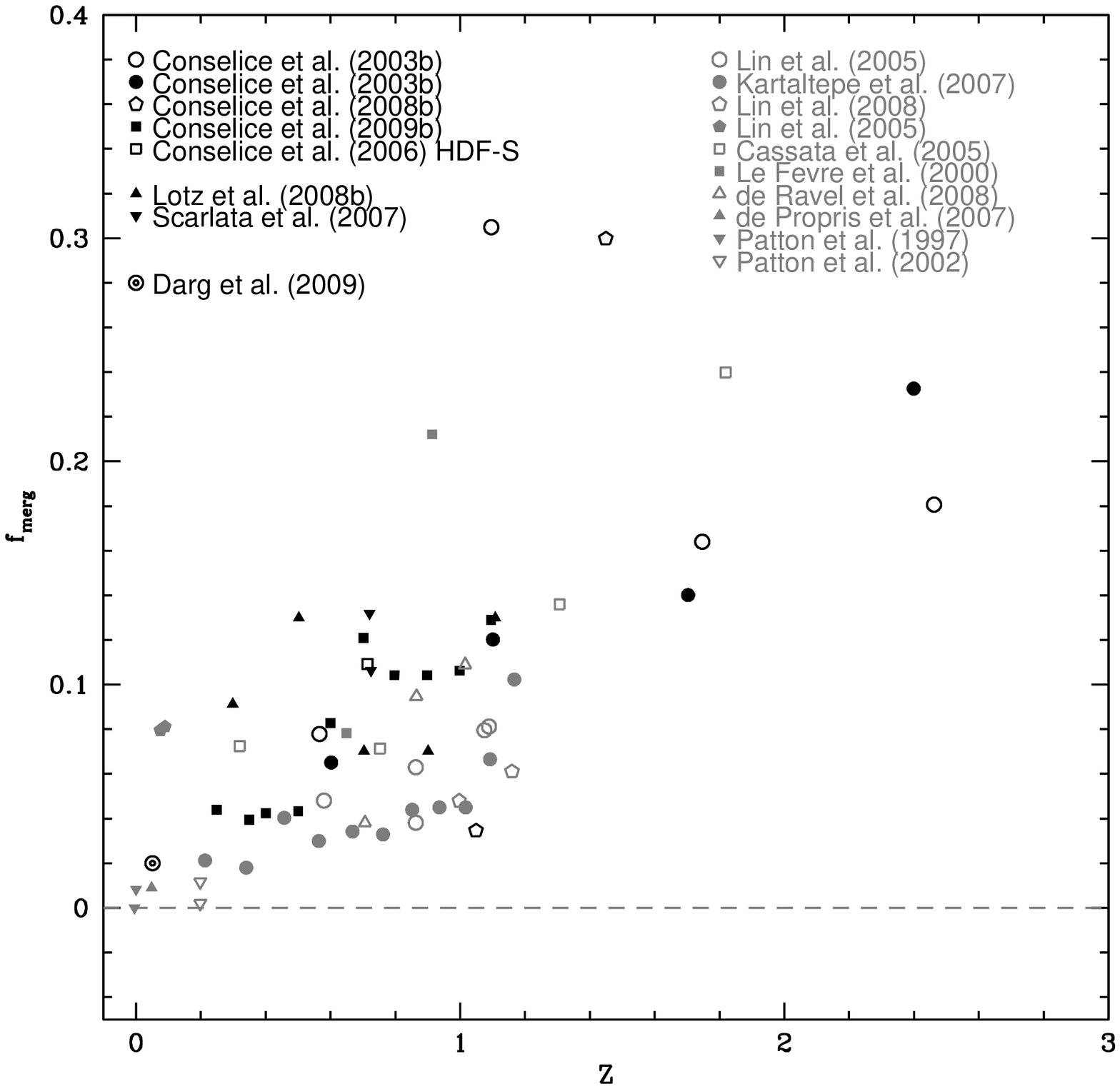}
\caption{\label{f:frac} The merger fraction ($f_{merg}$) as a function of redshift (z). The black points are based on quantified morphology estimates, the gray points are based on galaxy pair counts. The Conselice et al papers are based purely on the CAS classification system, \protect\cite{Lotz08b} uses the Gini/$M_{20}$ classification and \protect\cite{Scarlata07} both the CAS as the Gini and $M_{20}$ parameters. 
The results by \protect\cite{de-Propris07} and \protect\cite{Darg09} are hybrid approaches; \protect\cite{de-Propris07} looks at pair statistics but also galaxy asymmetry and \protect\cite{Darg09} used the visual identification of a merging pair in the GalaxyZOO project \protect\cite{galaxyzoo}. See for the data-sets used and the references Table \ref{t:cites}.}
\end{figure}

Because the volume probed increases with redshift, there were until recently --paradoxically-- better measures of the interaction fraction for higher redshift than for the local Universe. The SDSS search for close (and disturbed looking) pairs of galaxies \citep{Darg09} added the valuable local Universe interaction fraction, improving on the estimate by \cite{Patton97}, from close pairs in the Uppsala General Catalog \citep{Nilson73}. A compilation of merger fractions determined as a function of redshift is shown in Figure \ref{f:frac} and Table \ref{t:cites}. 
The scatter in the merger fractions, even for those determined over the same data is striking and it is tied to the definition of what constitutes a pair or disturbed morphology \citep[see for instance the discussion in ][]{Genel09}. The definition of morphologically disturbed became quantified in several schemes of morphological parameterisation schemes \citep[e.g.,][]{CAS,Lotz04}. 
Observational uncertainties are the time scale a merger is identifiable as one, the completeness of the various samples for each technique and the volumes considered. 
Similarly, a current substantial theoretical effort is to map the dark matter halo merger rates onto actual observable galaxy mergers \citep[][and reference therein]{Hopkins10}.

\begin{table}
\caption{The reference, data-set, and method for merger fractions in the local and distant Universe. }
\begin{center}
\begin{tabular}{l l l}
Reference			& Data-set 		& Criteria\\
\hline
\hline
Morphology			&				& \\
\hline
\cite{Conselice03b}		& HDF-N			& CAS \\	
\cite{Conselice08b}		& HUDF			& CAS \\
\cite{Conselice09b}		& EGS, COSMOS	& CAS \\
 \cite{Conselice05a}		& HDF-S				& CAS \\

\cite{Lotz08b}			& EGS			& G/$M_{20}$\\
\cite{Scarlata07}		& COSMOS		& CAS+G/$M_{20}$\\
\hline
\hline
Pair statistics:			&				& \\
\hline
\cite{Lin04}			& DEEP2			&\\
\cite{Lin08} 			& DEEP2			&\\
\cite{Kartaltepe07}		& COSMOS		&\\
\cite{de-Ravel09}		& VLT/DEEP		&\\
\cite{Cassata05}		& GOODS			&\\
\cite{Le-Fevre00}		& CFRS, HST		&\\
\cite{Patton97}			& CNOC1			&\\
\cite{Patton02}			& CNOC2			&\\
\cite{de-Propris07}		& MGC 			&\\
\hline
\end{tabular}
\end{center}
\label{t:cites}
\end{table}%

The morphological studies are largely based on optical, mostly B-band and restframe UV data. The reasoning goes that mergers trigger star-formation and the resulting increased surface brightness make the disturbed morphology easier to pick up \citep[although the increase in star-formation is not a given, see][]{Robaina09}. 
However, with the emergence of new and refurbished radio observatories in preparation for the future Square Kilometre Array \citep[SKA;][]{ska}, a new window on merger rates over cosmic times will be opening up: the 21 cm emission line of atomic hydrogen gas (\hi). The two SKA precursors, the South African Karoo Array Telescope \citep[MeerKAT;][]{MeerKAT,meerkat1,meerkat2}, and the Australian SKA Pathfinder \citep[ASKAP;][]{askap2, askap1, ASKAP, askap3,askap4} stand poised to observe a large number of Southern Hemisphere galaxies in \hi \ in the nearby Universe (z$<$0.2). In addition, the Extended Very Large Array \citep[EVLA;][]{evla} and the APERture Tile In Focus instrument \citep[APERTIF;][]{apertif,apertif2} on the Westerbork Synthesis Radio Telescope (WSRT) will do the same for the Northern Hemisphere. 
The advantage of \hi \ observations is that it contains both morphological and kinematic information of spiral disks. There is ample anecdotal evidence of disturbed \hi \ morphology during a merger \citep[see the compilation in][]{Hibbard01}\footnote{The \hi \ Rogues gallery: \\ \url{http://www.nrao.edu/astrores/HIrogues/}}
In this series of papers, we explore primarily the signature of gravitational interaction on the morphology of the (face-on) \hi \ disk. This is a suitable complement to any kinematic signature, which will be most clear in edge-on disks. Our motivation to move to the \hi \ perspective is that (a) the gas will be disturbed before the stellar disk is, (b) the \hi \ morphology will be more sensitive to minor interactions, which may dominate the number of interactions and (c) the \hi \ morphology will be intrinsically sensitive to gas-rich interactions. Minor and gas-rich interactions are expected to dominate at higher redshift, which makes an \hi \ perspective at low redshift a good local comparison.

In the previous papers in this series, we compared the \hi \ morphology to that in other wavelengths \citep{Holwerdapra09,Holwerda10b} and found it to be at least as good a tracer of mergers as any other wavelength. We defined an \hi \ parameter space to identify interacting galaxies \citep{Holwerda10c} and derived a time-scale for interactions to reside in this parameter space \citep[this paper's companion,][]{Holwerda10d}.
In this paper, the aim is to combine the morphological identification of mergers with the time-scales into a merger fraction and rate for the WHISP sample. The organisation of this paper is as follows: in section \ref{s:morph} we briefly describe the morphological parameters and selection criteria, in section \ref{s:limits} we discuss the limitations and applicability of these in the context of \hi data, in section \ref{s:whisp}, sections \ref{s:bdata} and \ref{s:maps} describe the WHISP basic data and \hi \ column density maps. In section \ref{s:vol} we derive the volume representative for the WHISP survey. In section \ref{s:frac}, we derive the merger fraction based on the number of pairs as well as the morphology and convert these into merger rates in section \ref{s:rate}. Sections \ref{s:disc} and \ref{s:concl} are our discussion and conclusions. 

\begin{figure}
\centering
\includegraphics[width=0.5\textwidth]{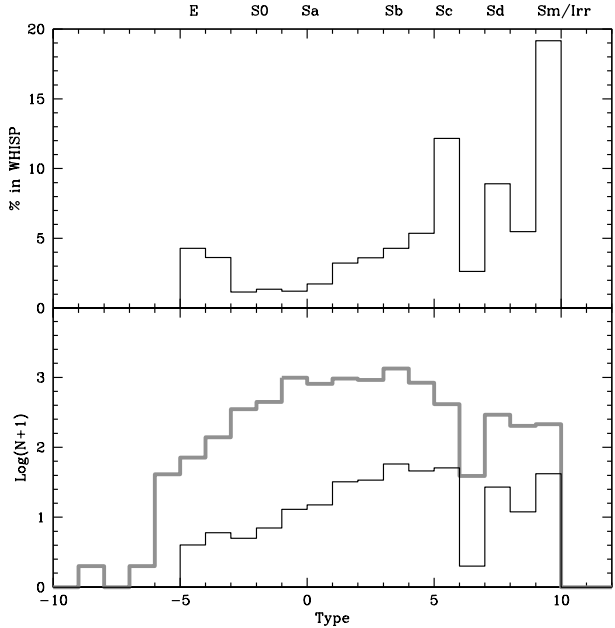}
\caption{\label{f:type} {\bf bottom panel}: The distribution of Hubble types in WHISP (black line) and the part of the UGC that conforms to the selection criteria for WHISP (thick gray line). {\bf top panel}: percentage of the UGC catalogue observed in WHISP as a function of type. There is a clear preference in the WHISP selection for later-type galaxies but Ellipticals are not specifically excluded and large early types would make the \hi \  flux cut. Hubble type determinations are from the 2MASS survey \citep{2MASS}, not the UGC.  }
\end{figure}

\section{Morphological Parameters and Merger Criteria}
\label{s:morph}

In this series we use the Concentration-Asymmetry-Smoothness parameters \citep[CAS][]{CAS}, combined with the Gini-$M_{20}$ parameters from \cite{Lotz04} and one addition of our own $G_M$. We discuss the definitions of these parameters in the previous papers, as well as how we estimate uncertainties for each. Briefly, given a set of $n$ pixels in each object, iterating over pixel $i$ with value $I_i$, position $x_i,y_i$ with the centre of the object at $x_c,y_c$ these parameters are defined as:

\begin{equation}
C = 5 ~ \log (r_{80} /  r_{20}),
\label{eq:c}
\end{equation}
\noindent with $r_{f}$ as the radial aperture, centered on $x_c,y_c$ containing percentage $f$ of the light of the galaxy \citep[see definitions of $r_f$ in][]{se,seman}.
\begin{equation}
A = {\Sigma_{i} | I_i - I_{180} |  \over \Sigma_{i} | I(i) |  },
\label{eq:a}
\end{equation}
\noindent where $I_{180}$ is the pixel at position $i$ in the galaxy's image, after it was rotated $180^\circ$ around the centre of the galaxy.
\begin{equation}
S = {\Sigma_{i,j} | I(i,j) - I_{S}(i,j) | \over \Sigma_{i,j} | I(i,j) | },
\label{eq:s}
\end{equation}
\noindent where $I_{S}$ is pixel $i$ in a smoothed image. The type of smoothing has changed over the years. 
We chose a fixed 5" Gaussian smoothing kernel for simplicity. 

The Gini coefficient is defined as:
\begin{equation}
G = {1\over \bar{I} n (n-1)} \Sigma_i (2i - n - 1) I_i ,
\label{eq:g}
\end{equation}
\noindent where the list of $n$ pixels was first ordered according to value and $\bar{I}$ is the mean pixel value in the image. 
\begin{equation}
M_{20} = \ log \left( {\Sigma_i M_i  \over  M_{tot}}\right), ~ {\rm for} ~ \Sigma_i I_i < 0.2 I_{tot}, \\
\label{eq:m20}
\end{equation}
\noindent where $M_i$ is the second order moment of pixel $i$; $M_i = I_i \times [(x-x_c)^2 + (y-y_c)^2 ]$. $M_{tot}$ is the second order moment summed over all pixels in the object and $M_{20}$ is the relative contribution of the brightest 20\% of the pixels in the object. 
Instead of using the intensity of  pixel $i$, the Gini parameter can be defined using the second order moment:
\begin{equation}
G_M = {1\over \bar{M} n (n-1)} \Sigma_i (2i - n - 1) M_i ,
\label{eq:gm}
\end{equation}

These parameters trace different structural characteristics of a galaxy's image but these do not span an orthogonal parameter space \citep[see the discussion in][]{Scarlata07}.
Originally, the above parameters were envisaged to classify the morphologies of galaxies 
but it was soon realized that a sub-space of the parameters is occupied by gravitationally interacting
late-types. 
\cite{CAS} and \cite{Lotz04} introduced several different criteria for the selection of merging 
systems in terms of the CAS and Gini-$M_{20}$ parameters
For optical data, \cite{CAS} define the following criterion:
\begin{equation}
A > 0.38, 
\label{eq:lcrit1}
\end{equation}
\noindent with some authors requiring A $>$ S as well. 

\cite{Lotz04} added two different criteria using Gini and $M_{20}$:
\begin{equation}
G > -0.115 \times M_{20} + 0.384
\label{eq:lcrit2}
\end{equation}
and
\begin{equation}
G > -0.4 \times A + 0.66 ~ \rm or ~ A > 0.4.
\label{eq:lcrit3}
\end{equation}
\noindent The latter being a refinement of the Conselice et al criterion in equation \ref{eq:lcrit1}.

These criteria were developed for optical morphologies, typically observed in restframe Johnson-B or SDSS-g. Therefore, in the third paper in this series \citep{Holwerda10c}, 
we defined several possible criteria specifically for the \hi \ perspective using the CAS-G/$M_{20}$-$G_M$ space of the WHISP survey \hi \ map sample.
We defined the Gini parameter of the second order moment, $G_M$ and a criterion for this parameter that selected most interacting galaxies:
\begin{equation}
G_M > 0.6,
\label{eq:crit1}
\end{equation}

Earlier in this series, we speculated that a combination of Asymmetry and $M_{20}$ could well be used to select interaction in \hi \ morphology in \citep{Holwerda10b}.
In  \cite{Holwerda10c}, we defined such a criterion as:
\begin{equation}
A > -0.2 \times M_{20} + 0.25.
\label{eq:crit2}
\end{equation}

Finally, we also defined one based on Concentration and $M_{20}$, following the example of the \cite{Lotz04} criteria (eq. \ref{eq:lcrit2} and \ref{eq:lcrit3}):
\begin{equation}
C > -5 \times M_{20} + 3. 
\label{eq:crit3}
\end{equation}
In \citep{Holwerda10b}, we found that this last criterion selected both the correct fraction of interacting galaxies and that it agreed most often with the previous visual identifications in the case of individual WHISP galaxies. We will  now explore merger rates based on the above criteria for \hi \ morphology.

\section{Limitations}
\label{s:limits}
Similar to other morphological selection schemes, we note that our approach is most sensitive to mergers involving at least one gas-rich late-type galaxy for the morphological selection and two in the case of pair selection of mergers. 
 \hi \  observations pre-select agains early-types (see Figure \ref{f:type} and section \ref{s:bdata}) and morphological disturbance is sensitive to unequal mass mergers \citep[cf][]{Lotz10b}. 
Therefore, this approach is complementary to existing morphological identification of mergers but dissimilar enough to warrant a separate estimate of time scales.

In \cite{Holwerda10d}, we compared the visibility time scales for the above criteria in the case of mergers of two equal mass spirals to those of a secularly evolving spiral. We find that the spiral-spiral merger is visible in morphological criteria during two stages before the final coalescence into the merger remnant; once during initial approach (before the stellar disk is disturbed) and during the second pass, before coalescence.
The total visibility time is approximately a Gigayear with some variance due to observation angle, different treatment of feedback from star-formation on the ISM in the simulation, and the relative gas fraction of the spiral disk. We note that the timescales for selection for merging and isolated (passively evolving) \hi \ disks become the same for resolutions coarser than the WHISP observations used here, e.g., the VLA Imaging of Virgo spirals in Atomic gas survey \citep[VIVA][]{Chung09}. A limitation to the simulations used \citep[originally from][]{Cox06a,Cox06b}, is that they are for Milky Way-size spiral galaxies only and do not consider minor mergers. 
Thus, since our approach may be sensitive to some minor merger scenarios, which possibly have much shorter visibility time-scales, the inferred visibility time from \cite{Holwerda10d} should be considered the upper limit for \hi \ morphological selection of mergers. 


 In a subsequent paper \citep{Holwerda10f}, we show that the \hi \ morphology is also sensitive to ram-pressure by a dense intergalactic medium but that one can select against ongoing or recent stripping with the Concentration index. For the WHISP survey we find in \cite{Holwerda10c} from a WHISP sub-sample and in \cite{Holwerda10d} from simulated \hi \ maps, that the level of contamination for the above parameters varies but is acceptable for large volume studies. For example, these still are noisy \hi \ maps in our morphological selection (Figure \ref{f:examples} and the Appendix in the {\em electronic version} of the journal).

\section{WHISP}
\label{s:whisp}

The dataset here are the observations done as part of the Westerbork \hi \ Survey of Irregular and Spiral Galaxies \citep[WHISP,][]{whisp,whisp2}. 
WHISP is a survey of the neutral hydrogen component in spiral and irregular galaxies with the Westerbork Synthesis Radio Telescope (WSRT). It has mapped the distribution and velocity structure of \hi \ in several hundreds of nearby galaxies, increasing the number of \hi \ observations of galaxies by an order of magnitude. The WHISP project provides a uniform database of datacubes, zeroth-order and velocity maps. Its focus has been on the structure of the dark matter halo as a function of Hubble type, the Tully-Fisher relation and the dark matter content of dwarf galaxies \citep{WHISPI,WHISPII, WHISPIII}. Until the large all-sky surveys with new instruments are completed, WHISP is the largest, publicly available data-set of resolved \hi \ observations.
We compiled a catalogue of basic data, obtained the highest available \hi \ column density maps and estimated the representative volume of WHISP.

\begin{figure}
\centering
\includegraphics[width=0.5\textwidth]{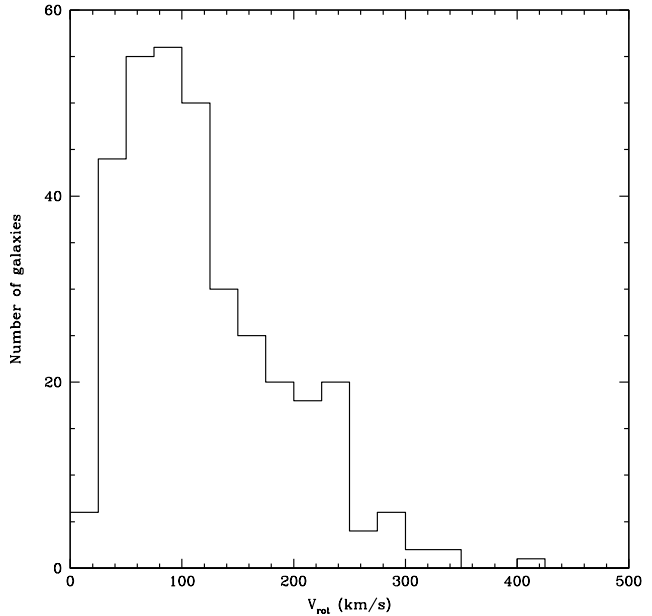}
\caption{\label{f:vrot} The distribution of rotation velocity ($v_{rot}$) over the complete WHISP sample.}
\end{figure}

\subsection{WHISP Basic Data}
\label{s:bdata}

Basic data for the WHISP sample came from the UGC catalogue, updated from HyperLEDA \citep{hyperleda}. 
We used updated positional data, preferring, in order, 2MASS \citep{2MASS}, the updated Uppsala Galaxy Catalogue 
positional data \citep{Cotton99}, the Principal Galaxy Catalogue \citep{PGC}, the original Uppsala Galaxy Catalogue 
positions \citep{Nilson73} and lastly the compilation of coordinates internal to HyperLEDA. 
The major and minor axis, came from the same catalogues in the same order. To define a sufficient sized area around the \hi \ disk, we multiplied the major axis with a factor seven. This is to speed up computation and leave out a galaxy's companions in the column density maps.

For the morphological information we again relied first on the 2MASS catalogue and secondly on the Uppsala Galaxy Catalogue and lastly on any information in HyperLEDA. The redshift information is primarily from \cite{Springob05} for many galaxies with the remaining ones filled in from a myriad of sources in HyperLEDA. Figure \ref{f:type} shows the distribution of Hubble types in the UGC and WHISP catalogue: there is a clear preference for late-types in WHISP.

\begin{figure}
\centering
\includegraphics[width=0.5\textwidth]{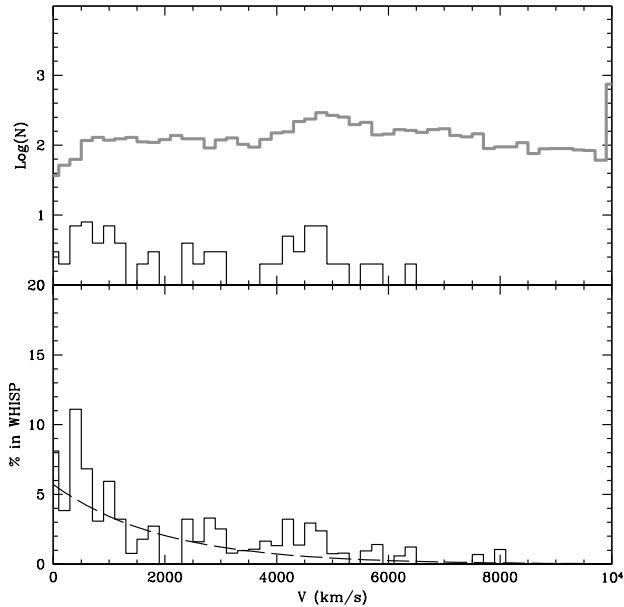}
\caption{\label{f:cz} {\bf top panel}: The distribution of recessional velocities of the WHISP galaxies and those in the part of the UGC that conforms to the selection criteria for WHISP. {\bf bottom panel}: percentage of the UGC catalogue observed in WHISP as a function of redshift. The fraction of WHISP galaxies drops off with redshift. The exponential fit to this drop-off is the dashed line, used to estimate the representative volume of the WHISP survey. }
\end{figure}

We also obtained HyperLEDA values for the rotational velocity ($v_{rot}$). Figure \ref{f:vrot} shows the distribution of $v_{rot}$ over the WHISP sample: WHISP selection prefers smaller ($v_{rot} < 120$ km/s), and more nearby systems (Figure \ref{f:cz}).

\subsection{WHISP Column Density Maps}
\label{s:maps}


The WHISP observation targets were selected from the Uppsala General Catalogue of Galaxies \citep{Nilson73}, with blue major axis diameters $> ~ 2\farcm0$, declination (B1950) $>$ 20 degrees and flux densities at 21-cm larger than 100 mJy, later lowered to 20 mJy. Observation times were typically 12 hours of integration. The galaxies satisfying these selection criteria generally have redshifts less than 20000 km/s ($z<0.07$).

The observational criteria (see above) are in effect a selection against early type galaxies (preferring spirals and irregulars, Figure \ref{f:type}), 
and a preference for galaxies below cz= 5000 km/s \citep[][chapter 2]{Noordermeer06}. Figure \ref{f:type} shows a histogram of the 
Hubble types in WHISP and the same volume in the UGC. There is a preference for later-type galaxies but no exclusive selection; 
only a few percent of the early types are selected and $\sim$ 10\% of the later types.

The WHISP data were retrieved from the ``Westerbork On the Web'' (WOW) project at ASTRON (\url{http://www.astron.nl/wow/}).  
We use the column density maps with the highest resolution available ($\sim$12" x 12"/sin($\delta$)). 

\subsection{The WHISP Volume}
\label{s:vol}

A definition of the WHISP volume is not straightforward as WHISP was not meant as a complete volume-limited sample of galaxies.
In the case of a {\em blind} \hi \ survey, the estimate of the volume sampled is complicated by the detection function of galaxies in 
the observations which depends on the bandwidth, frequency resolution and threshold used in the survey \citep[see][chapter 3]{Zwaan97,Zwaan00}. 
However, since WHISP is a {\em targeted} survey from an existing optical catalogue, we can compute the volume represented by 
the optical catalogue (UGC), estimate what fraction of the UGC the WHISP catalogue represents 
and thus what fraction of the UGC volume is representative of the WHISP survey.

Naively, the volume covered by the UGC with a declination over 20$^\circ$ and cz $<$ 5000 km/s (r=68.5 Mpc) is: 
$V_{UGC} (\delta>20^\circ) = (2 \pi/3)~ r^2 h =  (2 \pi/3)~ r^3 \times 1-sin(\delta) = 4.43 \times 10^5$ Mpc$^3$.
Of the \Nugc \ galaxies in the UGC, \Nwhisp \ galaxies are in WHISP; 4.17 \% of those in the volume.
However, to equate the WHISP volume to 4.17 \% of the UGC volume (18473 Mpc$^3$) would be simplistic as 
there is a bias towards nearby galaxies in the WHISP selection (see Figure \ref{f:cz}). 

Figure \ref{f:cz} shows the distribution in redshift of the WHISP sample and the total UGC sample ($\delta>20^\circ$), 
as well as the percentage of the UGC galaxies in WHISP. We fit an exponential distribution to the fraction and obtain 
the radial weighting function for the volume of the UGC corresponding to the WHISP sample: $w(r) = f_0 e^{-r/h}$
with  $f_0$ = 5.7 \% and h = 1947 km/s = 27 Mpc.  
To compute the WHISP volume we integrate over radius, weighting the radius with the above function: 
$V=  \int 2\pi r h w(r) dr = 6835$ Mpc$^3$, 1.5 \% of the UGC volume.
We will use this volume for our computation of volume merger rates further in this paper. 
Because the WHISP survey was never meant to be a volume-limited estimate, this estimate of the representative 
volume should be treated with caution. Fortunately, the future planned \hi \ surveys with ASKAP and APERTIF are 
set to be volume-limited.

	








\section{Merger Fraction}
\label{s:frac}

There are two ways for us to estimate the merger fraction of the WHISP sample: by counting the number of close pairs or to count the number of disturbed looking galaxies. 

We should note that in the lowest mass range ($M<10^{10} M_\odot$), the observed merger fractions are very high for redshift range z=0.2-1.2 \citep[$\propto 10$\%][]{Bridge07, Bridge10, Kartaltepe07, Lotz08b, Lin08, Conselice09b, Jogee09}.  Because the WHISP selection prefers nearby, irregular and smaller systems (Figures \ref{f:cz}, \ref{f:type} and \ref{f:vrot}), one can expect a high fraction of them to be merging.

\begin{table}
\caption{The galaxies in WHISP with one or more companions in the data-cube. Qualifiers of interaction (Int?) from either \protect\cite{Noordermeer05} (NM05) or \protect\cite{Swaters02} (SW02). }
\begin{center}
\begin{tabular}{l l l l }
Galaxy 		& Companions & Int? & Ref\\
\hline
\hline
 UGC 624  	& 2 			& y 	& NM05\\
 UGC 1437 	& 1 			& -	& -	\\
 UGC 2141 	& 1 			& n	& NM05\\
 UGC 2154 	& multiple  	& y 	& NM05\\
 UGC 2459 	& 1 			& -	& -	\\
 UGC 2487 	& 2 			& n	& NM05\\
 UGC 2916 	& 2			& y 	& NM05\\
 UGC 2941 	& 1 			& y 	& NM05\\
 UGC 2942 	& 1 (UGC 2943) 	& -	& -	\\
 UGC 3205 	& 3  			& n 	& NM05\\
 UGC 3382 	& 1 			& n 	& NM05\\
 UGC 3384 	& 1 			& -	& -	\\
 UGC 3407 	& 3 			& y 	& NM05\\
 UGC 3426 	& 1 			& y 	& NM05\\
 UGC 3546 	& 1 			& n 	& NM05\\
 UGC 3642 	& 1 			& y 	& NM05\\
 UGC 3698 	& 1 			& n	& SW02\\
 UGC 4458 	& 1 			& y 	& NM05\\ 
 UGC 4666 	& 1 			& n 	& NM05\\
 UGC 4806 	& multiple 	& -	& -	\\
 UGC 5060 	& 1 			& n 	& NM05\\
 UGC 5935 	& multiple  	& y	& SW02 \\
 UGC 6001 	& 1 			& n 	& NM05\\
 UGC 6787 	& 1 			& y 	& NM05\\ 
 UGC 7183 	& 1 			& -	& -	\\
 UGC 7353 	& 1 			& -	& -	\\
 UGC 7506 	& 1 			& n 	& NM05\\
 UGC 7989 	& 1 			& y 	& NM05\\ 
 UGC 8271 	& 3 			& y 	& NM05\\ 
 UGC 9642 	& 1 			& -	& -	\\
 UGC 9858 	& 1 			& -	& -	\\
 UGC 10791 	& 2 			& -	& -	\\
 UGC 11283 	& 1 			& -	& -	\\
 UGC 11951 	& 1 			& y 	& NM05\\ 
 UGC 12815 	& multiple  	& y 	& NM05\\ 
\hline
\end{tabular}
\end{center}
\label{t:frac}
\end{table}%

\subsection{Galaxy Pairs in WHISP}
\label{s:pairs}

There are several galaxies that have a close companion in the \hi \ datacube. Each datacube is a single WRST pointing (10' $\times$ 10') with a bandwidth of 320, 680, 1280 or 2560 km/s, depending on the velocity resolution used.  While this is not the typical selection criterion for pair selection \citep[see][]{Patton00}, we could use it as such since pairs are selected for proximity on the sky and in redshift. In the full WHISP catalog, there are 35 galaxies with one or more companions in the WHISP cube. Naively this translates to a close companion and hence merger fraction ($f_{mgr}$) of $\sim$ 10 \% of the WHISP sample. The merger fraction based on the close pairs depends on how many of those galaxies with companions one would consider merging. Typically, the velocity difference is taken to be less than 500 km/s to constitute a merging pair, so the datacube criteria are not stringent enough. If we go by the merger qualifiers from \cite{Swaters02} and \cite{Noordermeer05}, 10 of the 24 galaxies they classify and who have companions are not merging (68\% success rate, see table \ref{t:frac}). So the real merger fraction of the WHISP catalogue is closer to $\sim$7 \%, which puts it close to the local values from \cite{Patton97}, \cite{de-Propris07} and \cite{Darg09} (See Figure \ref{f:frac}) for the local volume.

\begin{figure*}
\centering
\includegraphics[width=\textwidth]{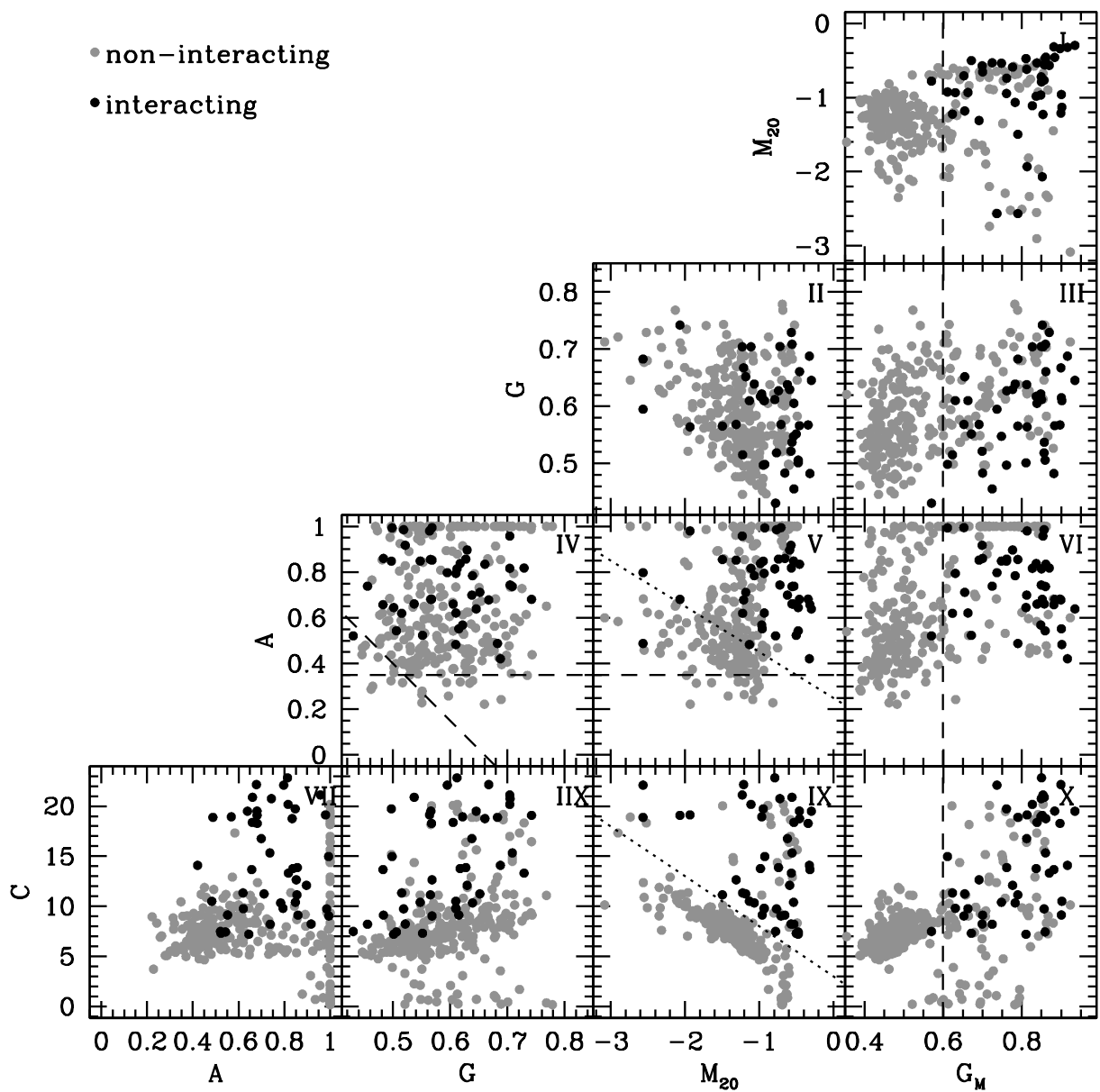}
\caption{\label{f:scarlata} The distribution of morphological parameters, Concentration (C), Asymmetry (A), Gini (G) and the contribution to the second order moment of the brightest 20\% of pixels ($M_{20}$), and the Gini coefficient of the second order moment of the pixels ($G_M$). 
Merger selection criteria from the literature are marked with dashed lines in panel II (equation \protect\ref{eq:lcrit2}), panel IV (equations \protect\ref{eq:lcrit1} and \ref{eq:lcrit3}), and V and VI (equation \protect\ref{eq:lcrit1}). Our selection criteria from \protect\cite{Holwerda10c} are marked with dotted lines; the $G_M$ criterion in panels I, III, VI and X (equation \protect\ref{eq:crit1}), the $A$-$M_{20}$ criterion in panel V (equation \protect\ref{eq:crit2}) and the C-$M_{20}$ criterion in panel IX (equation \protect\ref{eq:crit3}).
Those objects selected by this last criterion (additionally requiring that Asymmetry is not extreme; A!=1) are marked in the plot to illustrate. WHISP morphological values are in Table \ref{t:morph} in the electronic version of the manuscript }
\end{figure*}

\begin{figure*}
\centering
\includegraphics[width=0.32\textwidth]{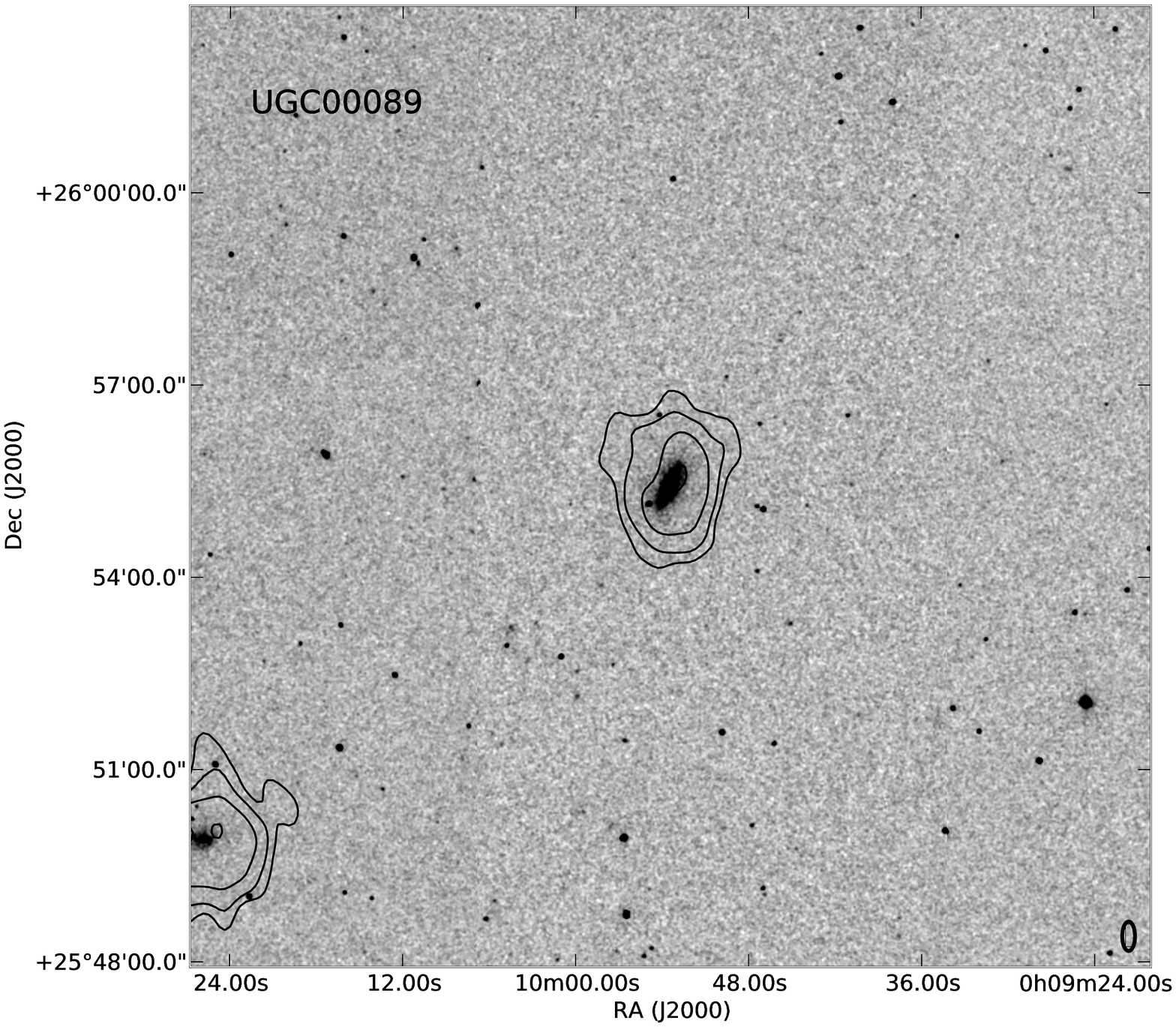}
\includegraphics[width=0.32\textwidth]{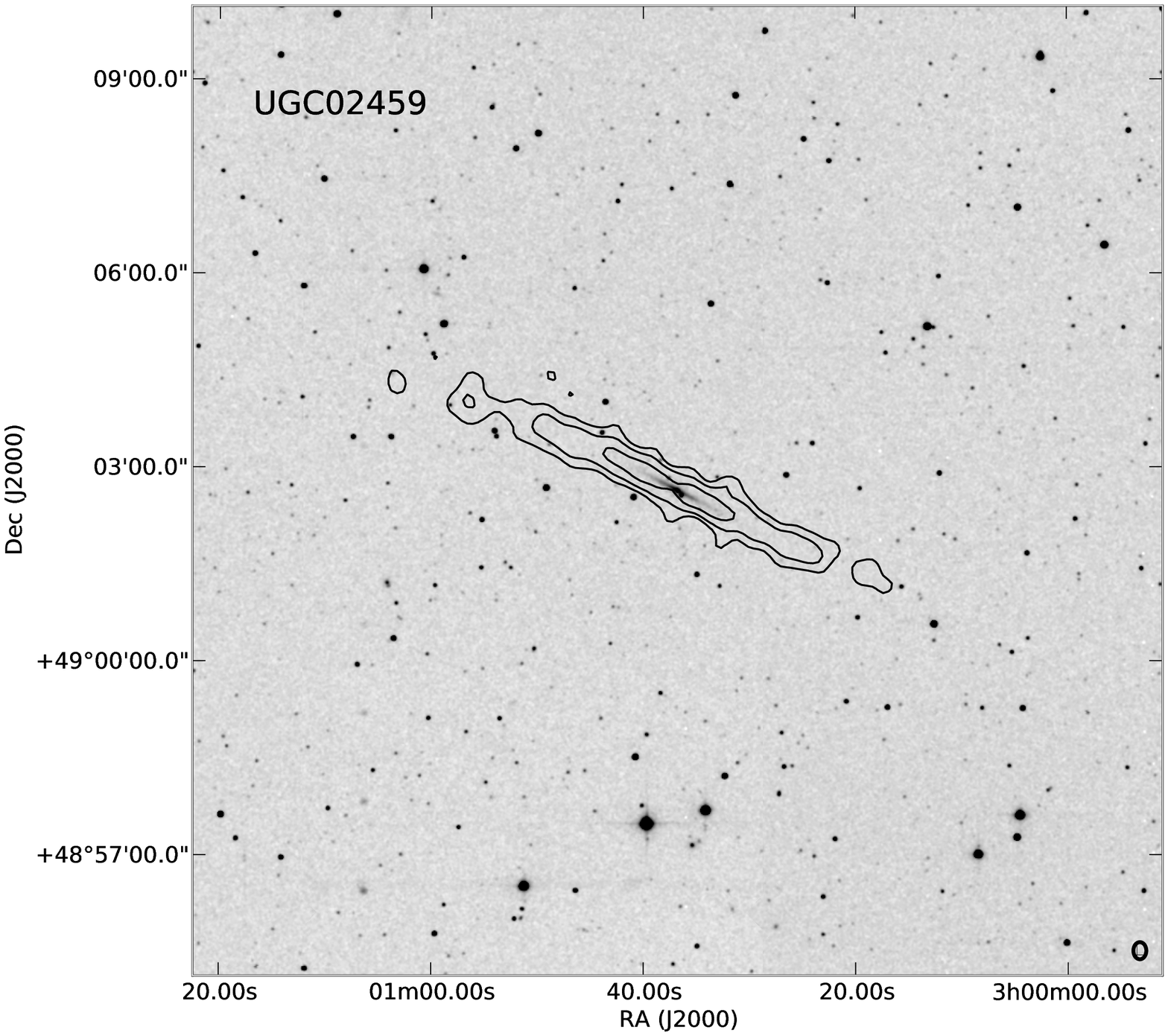}
\includegraphics[width=0.32\textwidth]{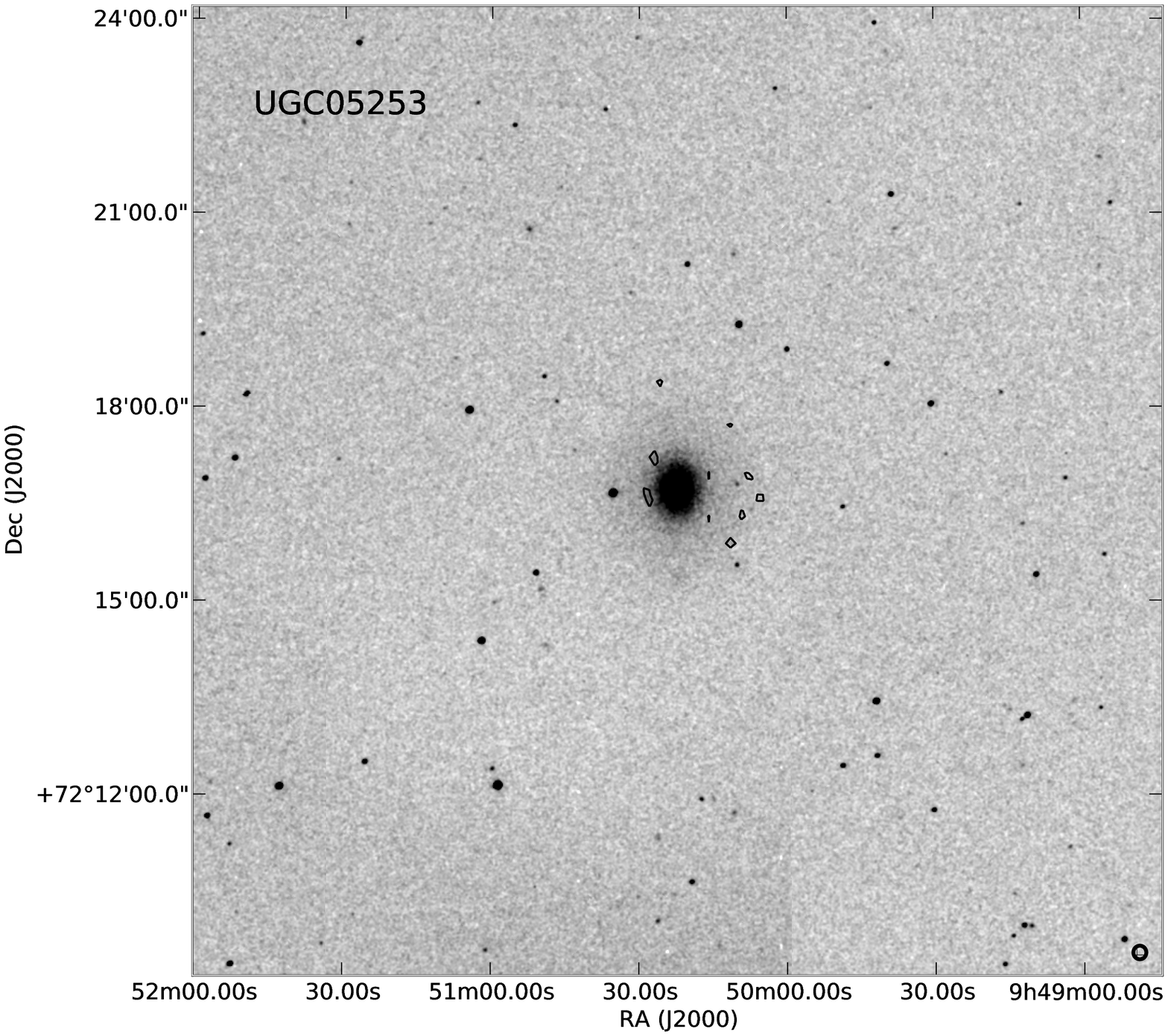}
\caption{\label{f:examples} Three random examples of the galaxies selected by the Concentration-$M_{20}$ criterion (eq. \ref{eq:crit3}). The grayscale image is the 2MASS K-band image and the contours are WHISP column density contours at 2.5, 5, 10 and $20 ~ \times ~ 10^{20}$ atoms/arcsec$^2$. . The full set of galaxies selected by this criterion is shown in the Appendix ({\em electronic version only}).}
\end{figure*}

\subsection{WHISP Merger Fraction from \hi \ Morphology}
\label{s:morph}

In \cite{Holwerda10c}, we identified the part of morphology parameter space that contains a representative number of the merging galaxies in a subsample of the WHISP database for which we had visual classifications of interaction using the \hi maps from either \citep{Swaters02} or \cite{Noordermeer05}. 
Based on a plot similar to Figure \ref{f:scarlata}, we concluded that criteria based on Asymmetry, $M_{20}$, Concentration and $G_M$ selected the correct fraction of interacting galaxies in a given sample (eq. \ref{eq:crit1} --\ref{eq:crit3}). 
Especially in the case of the Concentration-$M_{20}$ selection, we obtained not only the same fraction of interaction but this criterion agreed with the majority of the visual classification of the \hi \ map in individual cases. And in this paper's companion \citep{Holwerda10d}, we explored how long both these criteria and those from the literature (eq. \ref{eq:lcrit1}--\ref{eq:lcrit3}) select mergers in by their \hi \ morphology. 

We can now apply these selection criteria to the full WHISP sample. The values for the morphological parameters of the full WHISP sample are listed in Table \ref{t:morph} in the Appendix ({\em electronic version only}). Table \ref{t:mrg} summarized our results for morphological selection for each of the six criteria: the fraction of the total WHISP sample selected, the resulting volume density, the visibility timescale from \cite{Holwerda10d}, and the computed volume merger rates. For comparison, it also shows the values for the merger selection based on close pairs computed above. Figure \ref{f:scarlata} shows the parameter space highlighting those selected by the Concentration-$M_{20}$ criterion. We excluded those galaxies with A=1, as this extreme value is indicative of an incorrect central position ($x_c,y_c$) of the WHISP galaxy from HyperLEDA.
Starting with the best performing selection criterion (eq. \ref{eq:crit3}), we find 45 galaxies out of the 339 in the WHISP catalog are interacting, or 13 \%. The other selection criteria select much higher fractions. The next best performing criterion ($G_M$) selected mergers very cleanly in the N-body simulations but its timescale appears to be very resolution sensitive.
The Concentration-$M_{20}$ criterion selected these 45 galaxies based on their \hi \ morphology but we do not expect each to be a merger individually (see Figure \ref{f:examples} for some examples from the selection. All the \hi contours maps overplot on 2MASS-K images are shown in Figure A1 in the Appendix {\em electronic version}). 
Close to the selection criterion (dotted line in Figure \ref{f:scarlata}, panel IX), individual measures may scatter in and out of the selection. However, based on our experience in \cite{Holwerda10c}, the fraction of mergers in WHISP is correct and individual galaxies are likely to be interacting viewed in the \hi \ perspective. 
Often their optical appearance may be still undisturbed as these are in the earliest stages of the merger\footnote{See also the \hi \ Rogues gallery for many examples of disturbed \hi \ morphology in normal appearing spirals \url{http://www.nrao.edu/astrores/HIrogues/RoguesLiving.shtml}, \cite{Hibbard01}.}. From the morphologies of the selected galaxies (Figure \ref{f:examples} and the Apppendix in the electronic version of this paper), one can see that some noisy maps are still selected as well. A more uniform (pipeline-reduced) survey will suffer less from these spurious selections. 

The merger fraction we find from \hi \ morphology selection is higher than other authors find for the local universe; for example, \cite{Darg09} find 1-3\% of all galaxies in SDSS to be merging and \cite{Patton02} similarly find only a few percent  from galaxy pairs. However, our fraction is similar to those found at slightly higher redshifts (z$\sim$0.1, see Figure \ref{f:frac}). %
As pointed out in section 4, the \hi \ morphology selection is likely sensitive to some minor merger scenarios as well. Minor mergers are expected to dominate the number of ongoing mergers and could in part explain our higher fraction.

\begin{table*}
\caption{Interaction fractions, merger visibility time and merger rate for the WHISP sample based on different morphological selection criteria }
\begin{center}
\begin{tabular}{l | l l l l l}
Criterion							& Mergers	 & $f_{mrg}$	& $n_{mrg}$					& $T_{mrg}$	& $R_{mgr}$ \\
								& \#		& (\%)		& mergers Mpc$^{-3}$	& Gyr		& mergers Gyr$^{-1}$ Mpc$^{-3}$ \\
\hline
pairs								& 15		& 7			& 0.0035						& 0.5			& 0.007 \\		
\hline
$A > 0.38$                   				& 221  	& 65  	& 0.032 & 1.85    & 0.017 \\
$G > -0.115 \times M_{20} + 0.384$   	& 178  	& 53  	& 0.026  & 0       & \dots \\
$G > -0.4 \times A + 0.66 $       			& 235 	& 69   	& 0.034  & 0.15   & 0.23\\
$G_M > 0.6 $                  				& 81  	& 24  	& 0.012 & 0.80    & 0.015\\
$A < -0.2 \times M_{20} + 0.25 $     		& 151  	& 45  	& 0.022  & 0.9     & 0.025\\
$C > -5 \times M_{20} + 3 $          		& 45  	& 13 		& 0.0066  & 0.97    & 0.0068\\
\hline
\end{tabular}
\end{center}
\label{t:mrg}
\end{table*}%

\section{WHISP Volume Merger Rate}
\label{s:rate}

From the number of galaxies with a companion or the number of disturbed looking galaxies in a given volume ($n_c$ and $n_{dist}$ respectively), one can calculate the volume merger rate ($\rm R_{mgr}$), provided one has an estimate of the merger time scale ($\rm T_{mgr}$), the merger rate from pairs; $\rm R_{mgr}(pairs) = n_{c}/T_{mgr}(pairs)$ or the merger rate from morphology $\rm R_{mgr}(morph) =  n_{dist}/T_{mgr}(morph)$. From \cite{Holwerda10d}, we have an estimate of the {\it mean} merger time scale with some variance due to 
differences in merger conditions (type of feedback physics in interstellar matter, type of encounter, and gas masses of the disks) and perspective (face-on versus edge-on). 
\cite{Lotz10a, Lotz10b} note similarly that time scales depend on mass ratio and gas fraction for optical morphological selection.
 
Mergers were on average visible for 40\% of the 2.5 Gyr of the merger simulation, making our typical time scale $\rm T_{mgr}(morph) \sim 1$ Gyr.\citep[see][this paper's companion paper.]{Holwerda10d}, very similar to those used in the literature for morphological selection.  \cite{Patton00, Patton08}  use a merger time scale for pairs of $\rm T_{mgr}(pairs) = 0.5$ Gyr. The volume represented by the WHISP sample was computed above (\S \ref{s:vol}) as 6835 Mpc$^3$. 

Following our simple merger fraction of 7\% from the number of WHISP galaxies with companions in the datacube, we obtain a volume merger rate of $\rm R_{mgr}$(pairs) = $0.7 \times 10^{-2}$ mergers Mpc$^{-3}$ Gyr$^{-1}$. 
%
Merger rates based on \hi \ morphology can use a variety of selection criteria (Table \ref{t:mrg}) and the merger rate from the Concentration-$M_{20}$ criterion is $\rm R_{mgr}$(morph) = $6.8 \times 10^{-3}$ mergers Mpc$^{-3}$ Gyr$^{-1}$.

In comparison to the fraction found from galaxy pairs (7\%) with the number found from morphology (13\%), both seem to be in reasonable agreement with the morphological selection on the high side. This is in line with the discrepancy found between pair selection and morphological selection as can be seen in Figure \ref{f:frac}.
After factoring in the relative timescales a merger is visible as a close pair \citep[0.5 Gyr][]{Patton00, Patton08} or above the Concentraton-$M_{20}$ criterion \citep[$\sim $1 Gyr; ][]{Holwerda10d}, the volume merger rates agree very well: 7 and $6.8 \times 10^{-3}$ mergers Gyr$^{-1}$ Mpc$^{-3}$ for paired and morphologically disturbed \hi \ disks respectively.

\section{Discussion}
\label{s:disc}

The WHISP sample represents only a very small volume of the Universe and the resulting merger fraction and rates are uncertain as a result of that.
However, the reasonable consistency with much larger samples such as the Sloan Digital Sky Survey \citep[e.g.,][see Figure \ref{f:frac}]{Darg09} are cause for optimism as \hi \ morphology as a tracer of the merger fraction and rate of galaxies. 

Volume merger rates in the literature for the local Universe vary somewhat with sample and survey. 
\cite{Masjedi06} finds for luminous red galaxies in SDSS a volume merger rate of$R_{mgv}$ = $0.6 \times 10^{4}$ Gpc$^{-3}$ Gyr$^{-1}$ = $1.7 \times 10^{-4}$  h$^3$ Mpc$^{-3}$ Gyr$^{-1}$. 
\cite{DePropris07} found for galaxies of all types in the Millennium Galaxy Catalogue (MGC) a volume merger rate of $R_{mgv}$ = $5.2 \pm 1.0 \times 10^{-4}$ h$^3$ Mpc$^{-3}$ Gyr$^{-1}$,
and \cite{Patton08} find a volume merger rate for all galaxy types based on SDSS and MGC of $R_{mgv}$ = $1.4 \pm 0.1 \times 10^{-4}$ h$^3$ Mpc$^{-3}$ Gyr$^{-1}$ for {\em major} mergers.
In contrast, we find a volume merger rate of $\rm R_{mgr}$(morph) = $2 \times 10^{-3}$ mergers h$^3$ Mpc$^{-3}$ Gyr$^{-1}$ (h = 0.73 or $H_0$=73 km/s/Mpc), an order of magnitude more than those above. 
Since our merger fractions are similar within a factor two to those in the literature for the local Universe, the issue for the volume merger rate would either have to be the inferred WHISP volume, the timescale or a bias in the selection of galaxies. 

The merger time scale is unlikely to be the issue. The visibility time of the merger starts earlier in \hi \ than in the stellar perspective but it is not substantially different from what other authors have found. Substituting any other visibility time-scale from the literature for morphological or pair selection would not reduce the merger rate (selection times are typically less or equal to $\sim$ 1 Gyr). We are more sensitive to minor mergers and the implied shorter visibility time-scales but this is unlikely to be an order of magnitude effect. 

Alternatively, we may have to consider the possibility that the morphologically disturbed galaxies are not all gravitationally disturbed but may suffer from effects unique to the gas perspective, for example ram-pressure stripping affecting the appearance of the \hi \ disk. Compared to the observed fraction of mergers from other sources \citep[e.g.,][for the SDSS]{Darg09}, this is of order a factor two discrepancy. The agreement between volume merger rates from WHISP from the pairs and morphology contradict this however.

 The WHISP volume estimate is a likely source of the discrepancy as it is the most uncertain of our numbers. However, even with the naive, larger estimate of the WHISP volume ($18 \times 10^3$ Mpc$^3$), this is only a factor three, not an order of magnitude. 
 
The WHISP selection process favours late-type galaxies (Figure \ref{f:type}) and local small irregular galaxies (to complement the spirals) and it was never intended as a volume-limited sample. Hence an intentional or unintentional selection bias may well have been introduced. Mergers identifiable by their morphology are more likely to happen to the gas-rich late-types and the irregulars are confined to a local --smaller-- volume, and many of them will be tidally affected. The \hi \ perspective is likely to be more sensitive to unequal mass mergers as these can be identified much more readily (the contrast in gas surface densities is not as great as it is in stellar surface brightness). \cite{Lotz10b} points out how one expects a much higher merger fraction in lower-mass systems and \cite{Patton08} points out that a factor two discrepancy can easily be expected if lower mass systems are included in even a pair statistical analysis. In addition, \cite{Lotz10a} and \cite{Conselice09c} identify gas-rich mergers are the most easily identified by their morphology. 


Therefore, we suspect our result points to a higher merger fraction and volume merger rate for spirals and irregulars in the local Universe, and less to a gross error in our 
WHISP volume, merger fraction or the merger time-scale.
%
If this is the case, merger fractions and rates for at least these types may not evolve with redshift as dramatically as previously thought. 
However, to confirm this, one would need a volume-limited large \hi \ survey with sufficient resolution and sensitivity for both morphological selection as well as accurate pair identification.

The \hi \ perspective can be reliably used in the local Universe(z$\sim$0), where a spatial resolution can be achieved in large, all-sky surveys (e.g., WALLABY on ASKAP, Koribalski et al. {\em in prep.} and WNSHIS on WSRT, J\'ozsa et al. {in prep.}) but any morphological identification of interacting gas disks at higher redshift will have to wait for SKA. The resolution of the Pathfinder instruments (MeerKAT and ASKAP) may well be enough to identify close pairs in the proposed deep \hi surveys (DINGO on ASKAP, \cite{Meyer09}, Meyer et al. {\em in prep}, and LADUMA with MeerKAT, \cite{Holwerda10iska,Holwerda11aas}, Holwerda et al. {\em in prep.}). 

\section{Conclusions}
\label{s:concl}

In this paper, we explored the merger fraction and rate based on the \hi \ observations of the WHISP sample of galaxies. The sample is still small compared to other local references based on for instance the SDSS or the Millennium Galaxy Catalogue \citep{De-Propris05} but provides us with an indication how well the \hi \ surveys of the near future will perform in this respect. From the quantified morphologies of the WHISP column density maps , we conclude:

\begin{itemize}
\item [1.] The merger fraction in the WHISP sample is 7\% based on pairs, and 13 \% based on disturbed morphology. These percentages are consistent if one takes into account how long a merger is visible as a close pair of galaxies and how long as a morphologically disturbed \hi \ disk.
\item [2.] Assuming the representative volume of the WHISP sample is 6835 Mpc$^3$, and a merger visibility time scale of 1 Gyr, the merger rate for our selection criterion is $\rm R_{mgr}$(morph) = $6.8 \times 10^{-3}$ mergers Gyr$^{-1}$ Mpc$^{-3}$ in the local Universe, very close to the value of $\rm R_{mgr}$(pairs) = $7 \times 10^{-3}$ mergers Gyr$^{-1}$ Mpc$^{-3}$ for galaxy pairs in WHISP.
\item [3.] While the WHISP merger fractions and especially rates mutually agree, the merger rates are much higher than those reported in the literature. 
Selection effects in the WHISP survey, preferring dwarf and irregulars, rather than a gross error in the WHISP volume, could well account for the difference as well as variation in the quality of the WHISP maps across the sample.
Upcoming, volume-limited \hi \ surveys should provide an accurate measurement of the local merger rate from both \hi \ morphology as well as close pairs.
\end{itemize}

\section{Future Work}

The 21 cm window on the Universe is set to revolutionise our understanding of the merger rate of spiral and irregular galaxies as three independent measures of merging or gravitational interaction are available in the data: \hi \ morphology, kinematic signatures in \hi \ of interaction (e.g., lopsidedness of the profile, non-circular motions and an irregular velocity field) and the easy detection of physically close companions in the \hi \ data-cube. 

Planned \hi \ surveys with the SKA Pathfinder instruments, include an all-sky survey (WALLABY with ASKAP) and (WNSHIS with WSRT/APERTIF), and a medium deep survey to $z\sim0.4$  (DINGO) with ASKAP, and an extremely deep survey (LADUMA) with MeerKAT. 
Combined, these will revolutionise the volume probed with \hi \ and help shed light on the merger fraction using all three tracers; morphological, dynamic and pair identification in the local Universe surveys, and dynamical and close pairs identification our to higher redshift.

\section*{Acknowledgments}

The authors would like to thank the anonymous referee for his or her diligence, effort and inquiry, which resulted in a great improvement of the manuscript.
We acknowledge support from the National Research Foundation of South Africa. The work of B.W. Holwerda  and W.J.G. de Blok is based upon research supported by the South African Research Chairs Initiative of the Department of Science and Technology and the National Research Foundation. A. Bouchard acknowledges the financial support from the South African Square Kilometre Array Project. We acknowledge the usage of the HyperLeda database \url{http://leda.univ-lyon1.fr}. We have made use of the WSRT on the Web Archive. The Westerbork Synthesis Radio Telescope is operated by the Netherlands Institute for Radio Astronomy ASTRON, with support of NWO.

\end{document}